# Measurement-Based Quantum Computation

Tzu-Chieh Wei
C. N. Yang Institute for Theoretical Physics and Department of Physics and Astronomy, State University of New York at Stony Brook, Stony Brook, New York 11794-3840, USA

## Article Summary

Measurement-based quantum computation is a framework of quantum computation, where entanglement is used as a resource and local measurements on qubits are used to drive the computation. It originates from the one-way quantum computer of Raussendorf and Briegel, who introduced the so-called cluster state as the underlying entangled resource state and showed that any quantum circuit could be executed by performing only local measurement on individual qubits. The randomness in the measurement outcomes can be dealt with by adapting future measurement axes so that computation is deterministic. Subsequent works have expanded the discussions of the measurement-based quantum computation to various subjects, including the quantification of entanglement for such a measurement-based scheme, the search for other resource states beyond cluster states and computational phases of matter. In addition, the measurement-based framework also provides useful connections to the emergence of time ordering, computational complexity and classical spin models, blind quantum computation, etc. and has given an alternative, resource-efficient approach to implement the original linear-optic quantum computation of Knill, Laflamme and Milburn. Cluster states and a few other resource states have been created experimentally in various physical systems and the measurement-based approach offers a potential alternative to the standard circuit approach to realize a practical quantum computer.

## Keywords

## Introduction

Quantum computation has the potential to perform specific computational tasks more efficiently and hence faster than current classical computers (Nielsen & Chuang, 2002). Over the past decade, a few small-scale quantum computers, whose size ranges from a few to about seventy quantum bits (qubits), have been built and put into action. The technology has become increasingly mature and it is likely that quantum computers will soon perform computational tasks beyond what current classical computers can efficiently simulate (Arute et al., 2019).

A natural framework for quantum computation is the standard circuit model, where an array of qubits are appropriately initialized, such as in the logical 0 state, and depending on the algorithmic task, a sequence of quantum gates (typically one-qubit and two-qubit) are applied to the array of qubits; finally, readout is done by measuring individual qubits in the logical 0/1 basis, the so-called computational basis. In addition to the circuit model, the adiabatic quantum computational model does not use gates but rather time-dependent, smoothly or adiabatically varied Hamiltonians (Averin, 1998; Farhi et al., 2000; Kadowaki & Nishimori, 1998). Both rely on the unitary property of either quantum gates or Hamiltonian evolution.

In contrast, measurement-based quantum computation, which originated from the work of Raussendorf and Briegel on the one-way quantum computer (Raussendorf & Briegel, 2001) utilizes local measurement to drive computation. Measurement is often regarded as a mechanism that destroys coherence in quantum states. The key feature to understand how measurement can achieve unitary operation is entanglement. The broader measurement-based framework is currently being explored as an alternative approach to realize a quantum computer.

## Part one: Quantum computation and measurement-based approaches

### Early development in quantum computation

The earliest notion of quantum computation goes back to the early 1980s. Paul Benioff published a paper that described a microscopic model of the classical Turing machine using quantum mechanics (Benioff, 1980). Yuri Manin also suggested the idea of quantum computation in his book "Computable and Noncomputable" (Manin, 1980). In a conference in 1981, Feynman discussed "Simulating Physics with Computers" and pointed out that it was not possible to simulate a quantum system efficiently with classical computers (Feynman, 1982). Therefore it was natural to consider simulating a quantum system with another quantum system that is well-controlled or, in other words, with a quantum-mechanical computer (Feynman, 1985). The most prominent work that suggests the potential quantum advantage is described in a paper by Shor that showed a quantum computer could, in principle, factorize a large integer number almost exponentially faster than any currently existing classical methods (Shor, 1994). To get a sense of the time complexity, if it takes 1 second to factor a 30-digit number for both classical and quantum computers, then it takes about 3 years for the classical computer to factorize a 100-digit number, but about 40 seconds for a quantum computer. To factorize a 300-digit number will take about a third of the age of the universe for a classical computer but only about 10 minutes for a quantum computer. At present such a powerful quantum computer does not exist. However, the potential capability prompted a great interest in both theoretical and experimental quantum computation and information science. The progress of quantum technology in the past few decades shows promising advances towards these goals.

### Rules of quantum mechanics and the circuit model of quantum computation

To understand how quantum computation works, it is essential to understand the governing rules stemming from quantum mechanics. Three of them are particularly important: (1) Superposition, (2) Evolution, and (3) Measurement. For an explanation of these rules, see e.g., (Susskind & Friedman, 2014). Superposition appears in classical waves, and in quantum mechanics, it allows quantum states, like vectors, to add or interfere. In fact, by representing quantum states as vectors (such as single-qubit states indicated by arrows in the `Bloch' sphere; see Figure 1(a)), how they evolve in time is governed by Schroedinger's equation, whose effect is to apply a suitable unitary matrix to the vector representing the quantum state. A quantum gate is built from the action of evolution. For example, the goal of the so-called NOT gate is to flip the arrow pointing to the north pole to the south pole in the Bloch sphere and vice versa; see Fig. 1(b). To so do, the evolution may begin with the north pole and follow the path of a meridian to the south pole. Another example is the so-called Hadamard gate, which consists of two steps (see Figure 1(b)): (1) rotation around the y-axis by -90° and followed by (2) rotation around the z-axis by 180°. The effect is to rotate $|0\rangle$ to $(|0\rangle + |1\rangle)/\sqrt{2}$, and $|1\rangle$ to $(|0\rangle - |1\rangle)/\sqrt{2}$. By using a sequence of three Euler rotations, an arbitrary one-qubit state $\alpha|0\rangle + \beta|1\rangle$ can be arrived at from $|0\rangle$, where α and β are two complex numbers that satisfy $|\alpha|^2 + |\beta|^2 = 1$.

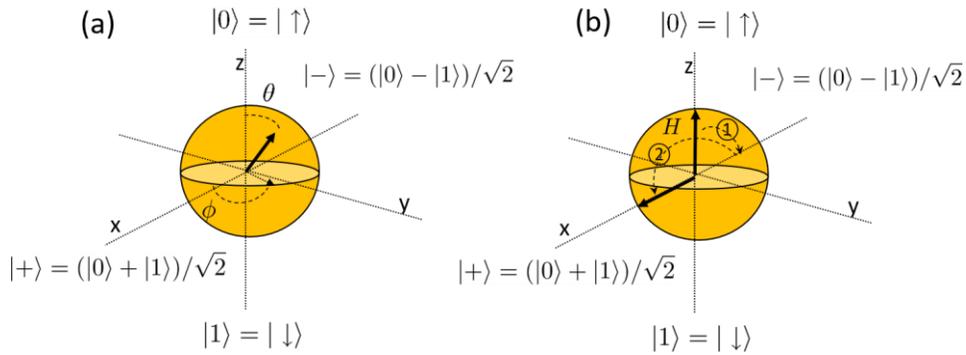

*Figure 1. Bloch sphere. (a) A vector represents the state of one qubit. (b) Illustration of the Hadamard gate acting on a vector pointing to the north pole. It first rotates it -90 degrees around the y-axis and then 180 degrees around the z-axis.*

The evolution under Schroedinger's equation is deterministic; in contrast, measurement of a quantum state generally yields random outcomes, and the distribution of outcomes also depends on the basis or the axis of the measurement. The rule of measurement in quantum mechanics states that the act of measuring an observable $O$ projects the system to an eigenstate of $O$, and the observed value is the associated eigenvalue. In the case of one qubit, the observable $O$ defines an axis cutting through the center of the Bloch sphere, and the two intersecting poles are the two possible outcomes. Unless the arrow representing the quantum state aligns exactly with one of the poles, the measurement outcome appears randomly and the outcome corresponding to either pole can appear. The probability distribution governing the random outcomes obeys the so-called Born rule, given by the modulus square of the coefficient of that eigenstate in the quantum state to be measured, and depends on the relative orientation of the state vector with the measurement axis.

The usual measurement result of 0 and 1 is represented as the axis connecting the north and south poles on the Bloch sphere. But measurement along the x axis that intersects the equator gives rise to two possible outcomes corresponding to $|+\rangle = (|0\rangle + |1\rangle)/\sqrt{2}$ and $|-\rangle = (|0\rangle - |1\rangle)/\sqrt{2}$. Practically, such a measurement can be achieved by carrying out the typical energy eigenbasis (Z) measurement in the 0 and 1 basis after the Hadamard rotation to induce the basis change (from X to Z or vice versa).

A quantum computer has many qubits, and there are an exponential number, $2^N$, of basis states for $N$ qubits, ranging from $|0 \ldots 0\rangle$ to $|1 \ldots 1\rangle$. Description of such a vector and its change in time requires an exponential number of complex numbers which is intuitively why quantum computers are difficult to simulate by classical computers.

Even with just two qubits, a natural consequence of quantum mechanics yields an exotic feature called entanglement, that appears in a quantum state of $(|00\rangle + |11\rangle)/\sqrt{2}$, which can be achieved by preparing the two qubits in $|00\rangle$ initially, applying the Hadamard gate to the first qubit (which rotates it from the north pole to a point on the equator: $|+\rangle = (|0\rangle + |1\rangle)/\sqrt{2}$), and then acting on them by a two-qubit CNOT gate (which flips the second bit only if the first is 1), just like the first two gates shown in Figure 2. The sequence takes $|00\rangle$ to $(|0\rangle + |1\rangle)|0\rangle/\sqrt{2}$ and then to $(|00\rangle + |11\rangle)/\sqrt{2}$.

A quantum computer, in a nutshell, implements a large unitary matrix $U$ on a vector of $2^N$ components representing $N$ quantum bits. A mathematical result (DiVincenzo, 1995) shows that any such unitary matrix $U$ can be decomposed into a sequence of one- and two-qubit gates, where one-qubit gates are simply performing local rotations and two-qubit gates are generating entanglement. The CNOT gate is the only two-qubit gate that is needed (Barenco et al., 1995); other entangling gates, such as a Controlled-Z gate may also be used instead. Such one- and two-qubit gates form the universal set of gates (DiVincenzo, 1995); such a notion of universality already exists in classical computation, with the set of AND, OR, and NOT gates being universal. From this picture of quantum computation, entanglement is created by quantum gates and subsequently reduced or destroyed by measurement. *In measurement-based quantum computation, the universal set of gates needs to be implemented by measurement.*

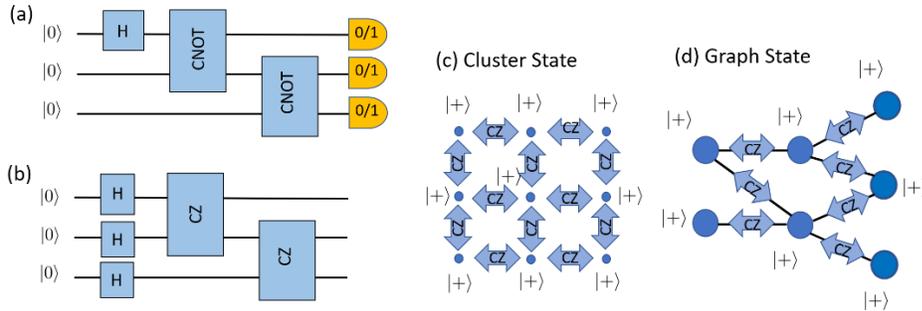

*Figure 2. Quantum circuits. (a) It is a quantum circuit of three qubits: first, a Hadamard gate is applied to the first qubit, transforming $|0\rangle$ to $|+\rangle$, then the CNOT gate is applied to the first and second qubits, followed by another CNOT gate acting on qubits 2 and 3. Each qubit is read out in the 0/1 basis. (b) A circuit to generate a one-dimensional three-qubit cluster state. After the three Hadamard gates, the three qubits become $|+\rangle$, and the pairwise CZ gates transform them into a chain in the cluster state. (c) A illustration of a two-dimensional **cluster state** in a 3-by-3 array of spins. This also serves as the definition of the 2d cluster state. (d) The cluster state can be generalized to any **graph state**, where pairwise CZ gates apply to a pair of qubits (initially in $|+\rangle$) according to the edges in the graph.*

*Table 1 Definitions of some terminology*

| Graph states: | Qubits reside on the vertices of a graph. The graph state can be defined by a procedure---all qubits are initialized in the $|+\rangle$ states and Controlled-Z gates are applied pairwise to a pair of qubits that share an edge. The resultant state is a graph state. See Fig. 2. |
|---|---|
| Cluster states: | A cluster state is a graph state when the underlying graph is a regular graph, such as a one-dimensional lattice or a two-dimensional square lattice. See Fig. 2. |
| Matrix product states: | A matrix product state is a quantum state whose coefficients in some expansion of basis states can be given via a product of matrices. This is usually used to describe one-dimensional quantum states. |
| Projected-entangled-pair states: | A projected-entangled-pair state is a quantum state that can be described by a projection of local virtual qubits or qudits to local physical degrees of freedom, and the virtual qubits are initially entangled pairwise with a neighboring virtual qubit. Matrix product states are special cases. See Fig. 7. |
| Tensor-network states: | A tensor-network state is a quantum state whose coefficients in some expansion of basis states can be given via a contraction of a tensor network. A tensor network is a collection of tensors located, e.g., at vertices of a graph. Edges connecting two vertices correspond to contraction, i.e., summing over identical indices. The local tensors are related to projections in the projected-entangled-pair states. They are in fact equivalent descriptions. Matrix product states are special cases. See Fig. 7. |
| Bell-state measurement: | This is also called Bell-basis measurement. It corresponds to a measurement on two qubits and the effect of the measurement is to project the two qubits to any of the four Bell states. See Fig. 4. |

**Measurement-Based Quantum Computation**

Besides the circuit model, there are other frameworks of quantum computation, such as adiabatic quantum computation, which are still based on the unitary evolution of a quantum system. Topological quantum computation utilizes the properties of the so-called anyons, which under exchange of pairs of anyons, i.e. braiding, can induce unitary transformation that can be used for quantum gates. The subject of interest here is measurement-based quantum computation (MBQC), which uses measurement to achieve emulation of unitary circuits. It originated from the pioneering work of Raussendorf and Briegel on the one-way quantum computer (Raussendorf & Briegel, 2001). Subsequent works resulted in some variants. The variants that will be discussed include the teleportation-based, state-transfer-based, and correlation-space approaches, which provide useful perspectives to appreciate the original one-way model and further development of the measurement-based quantum computation.

*One-way quantum computer and cluster states*

In around 2000, Raussendorf and Briegel showed that quantum computation was possible by merely performing individual single-qubit measurements, which they called the one-way quantum computer (Raussendorf & Briegel, 2001). The key necessary ingredient is the high persistent entanglement residing in the cluster state that Raussendorf and Briegel exploited (Briegel & Raussendorf, 2001). A cluster state can be described as follows. Qubits are sitting on the vertices of a graph, and the edges describe an Ising-like interaction (Ising, 1925) between two adjacent spins. The only nontrivial effect is to induce a sign change in the state $|11\rangle$, so that $|11\rangle$ becomes $-|11\rangle$ after the interaction. This is also called a Controlled-Phase or Controlled-Z (CZ) gate. If initially all the qubits are in $|+\rangle$ state, and the system after pairwise action of Controlled-Z gates will end up in a graph state. *The cluster state is a special graph state on a regular lattice*, such as the square lattice; see Fig.2 (c)&(d).

Single-qubit measurement can only decrease the amount of entanglement, and hence the computation via measuring qubit by qubit in the cluster is "one-way". Any quantum circuit in the standard circuit model can be translated to a measurement pattern on all the qubits of the cluster state. Execution of the measurement pattern with possible adaptation then drives computation and at the same time, the entanglement as a *resource* (for computation) is being `consumed'.

In more detail, in a two-dimensional square array of qubits initially in the cluster state, structures can be 'carved out' to form a backbone of computation by measuring unwanted qubits in the z-axis. Such a backbone mimics the structure of a quantum circuit. For a segment of five linear sites, any rotation in the Bloch sphere can be achieved by using a combination of three Euler angles (α,β,γ) shown in Figure 3(a). The symbols in the circles represent the angles of the measurement axes as measured from the positive x-axis on the x-y plane. Given that measurement gives random outcomes, to make the computation stay on track, subsequent measurement axes may need to be adapted, e.g., by flipping the angle with a minus sign. This adaptation is the feedforward that is needed to make the desired unitary gates deterministic (Raussendorf et al., 2003). To complete the universal gate set, a two-qubit entangling gate such as the Controlled-NOT gate is needed. One example to realize this is illustrated in a structure of 'I' shape junctions; see Figure 3(b). It is interesting to note that the adaptation of measurement axes is not necessary to implement the CNOT gate, in contrast to arbitrary one-qubit gates. There are other variants of these 'LEGO' pieces for quantum gates (Raussendorf et al., 2003; Raussendorf & Briegel, 2001). By placing these pieces on a 2D grid, any quantum circuits can be realized by local measurement. Hence, the 2D cluster state can be regarded as a universal resource for quantum computation.

The above explanation of the one-way quantum computer relies on the mapping of a quantum circuit to a measurement pattern in the cluster state. In fact, it is not necessary to use the circuit-simulation picture; instead an "intrinsic" one-way computer based on the consideration of measurement, time ordering, and deterministic computation can be used (Raussendorf et al., 2016; Raussendorf & Briegel, 2002).

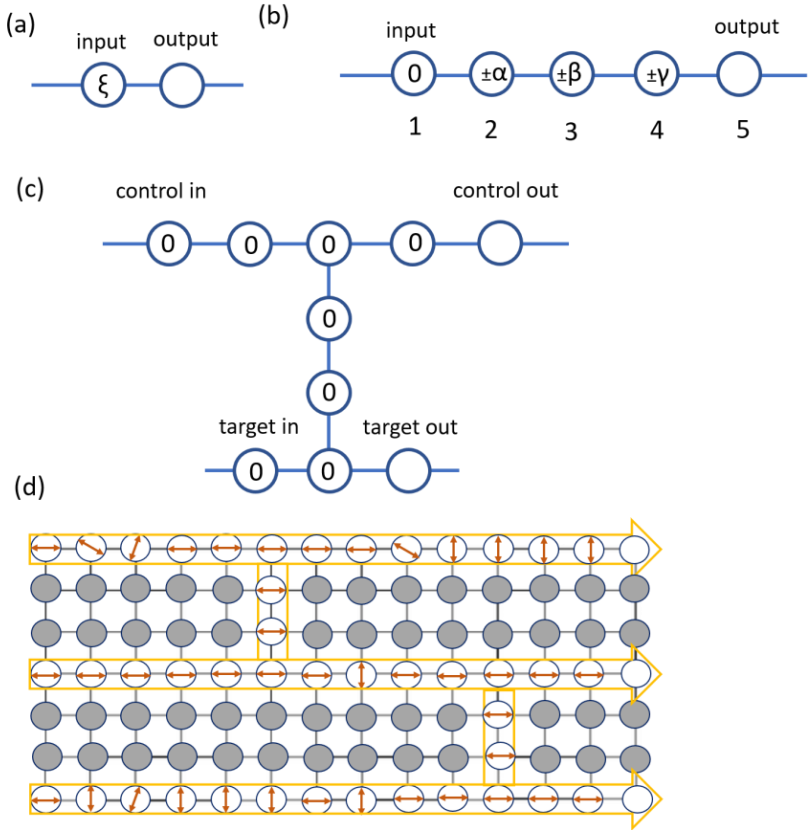

*Figure 3. Realizing universal gates. (a) Basic block of an input-output structure on two qubits which have entanglement formed by a CZ gate. The symbol ξ represents the basis of measurement, which is the observable $\hat{O}(\xi) \equiv \cos\xi\, \sigma_x + \sin\xi\, \sigma_y$. Denote the random measurement outcome by s=0 or 1, the input state $|\psi_{in}\rangle$ will be transformed to $|\psi_{out}\rangle = He^{i\xi\sigma_z/2}(\sigma_z)^s|\psi_{in}\rangle$, where tht output resides on the second qubit. (b) Entanglement structure for an arbitrary one-qubit gate. Specific gates may need fewer sites. The + or – sign inside the circles represent adaptation of measurement axis. Labeling the measurement outcomes by $s_i$, the signs of the measurement axis on qubits 2,3, and 4, are determined by $(-1)^{s_1}$, $(-1)^{s_2}$, and $(-1)^{s_1+s_3}$, respectively. This is the adaptation of later measurement axes, dependent on previous measurement outcomes, which requires feedforwarding the information. This also imposes a time ordering 1→2→3→4→5. The three angles α, β, and γ are related to the Euler angles that define a general rotation. (c) Structure for realizing the two-qubit CNOT gates. Interestingly, these measurements are all fixed at $\hat{O}(\xi=0)$, can all be performed at the same time step, and no feedforwarding is needed if the CNOT gate is the last operation. However, the measurement outcomes are needed to adapt the measurement axes for later gates. (d) Example of a 3-qubit circuit in the one-way quantum computer picture, realized in a grid of 14 x 7 qubits, initialized in the cluster state and consumed by measurement from left to right. Dark circles represent the measurement in the z-axis. Double-headed arrows in the circles illustrate the axes of measurement, i.e., the angles ξ. Note that the three Pauli matrices $\sigma_x$, $\sigma_y$, and $\sigma_z$ are also conveniently represented by X, Y, and Z, respectively. See also Ref. (Raussendorf & Briegel, 2001).*

## Other approaches of measurement-based quantum computation

Since the invention of the one-way quantum computer by Raussendorf and Briegel, there have been attempts to understand this novel approach of quantum computation by using different perspectives, including teleportation, state transfer and tensor network.

*Teleportation-based measurement scheme for quantum computation.* The teleportation-based construction of quantum gates was earlier proposed by Nielsen and Chuang (Nielsen & Chuang, 1997) and by Gottesman and Chuang (Gottesman & Chuang, 1999). The basic setup of teleportation is illustrated in Figure 4(a), where an unknown qubit state $|\psi\rangle$ can be transferred to a third qubit by using an entangled pair $(|00\rangle + |11\rangle)/\sqrt{2}$, one of the four Bell states, shared between a second and a third qubits and then joint measurement on the first two qubits. A correcting operation that depends on the measurement outcomes completes the teleportation and recovers the unknown state. By using such a teleportation-based approach (Bennett et al., 1993; Gottesman & Chuang, 1999; Nielsen & Chuang, 1997), Nielsen showed that it is possible to perform universal quantum computation using only measurements (needed to create entanglement that will be used to mediate gate operations) and quantum memory (needed to store quantum information and entanglement) (Nielsen, 2003), without the need of a prior entangled resource state. The key intuition is that by allowing measurement on two or more qubits, entanglement can be created. Nielsen generalized the quantum teleportation protocol by using a locally rotated Bell state (by $U$), and showed that a quantum state could be teleported so that the output state is acted by a random Pauli operator σ (associated with the usual teleportation) and additionally the desired gate $U$; see Figure 4(b). The random Pauli operator arises due to unpredictable measurement outcomes and can be probabilistically canceled by repeatedly performing the above "teleportation" procedure as in Figure 4(b) until the product of these Pauli operators cancels one another and becomes identity. By using a four-qubit state that was defined by rotating two pairs of Bell states by a two-qubit gate $U$, two-qubit gates can be achieved; see Figure 4(c). Such a four-qubit state can be created by a four-qubit measurement and it can be used to induce a two-qubit gate such as the CNOT gate in Nielsen's scheme. The upshot is that universal quantum computation can be done by a combination of two- and four-qubit measurement.

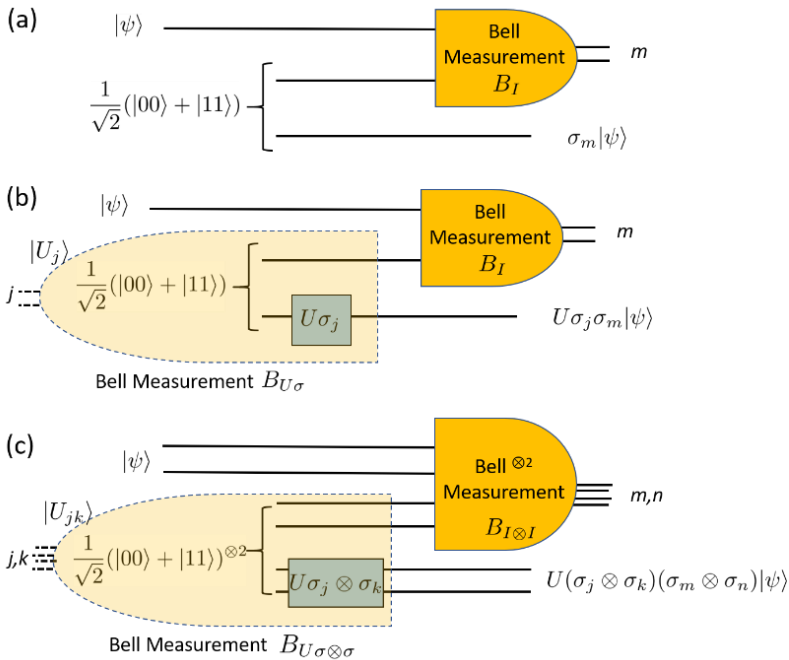

*Figure 4. Measurement-based quantum computation via teleportation. (a) The standard **teleportation**. It consumes a Bell state $(|00\rangle + |11\rangle)/\sqrt{2}$ and performs a Bell-basis measurement together on the known state ψ and one qubit of the Bell state, which results in transferring the quantum state $|\psi\rangle$ to the other unmeasured qubit of the Bell state, up to a random Pauli operator. The **Bell-basis measurement** projects a two-qubit quantum state to any of the four basis states: $|\Phi^\pm\rangle = (|00\rangle \pm |11\rangle)/\sqrt{2}$ and $|\Psi^\pm\rangle = (|01\rangle \pm |10\rangle)/\sqrt{2}$. The random Pauli operator arises because that the measurement is intrinsically random and the randomness cannot be removed. (b) Teleportation-based one-qubit gates. Different from (a), a gate $U\sigma_j$ is applied to the third qubit to rotate the Bell state and there is no correction operation on the third qubit. This circuit can be interpreted as using a rotated Bell state as a resource for teleportation. Such a rotated Bell state can be created probabilistically by performing an appropriate two-qubit measurement. (c) Teleportation-based two-qubit gates. Generalizing the consideration in (b) to two pairs*

*of Bell states rotated by a two-qubit gate gives an equivalent circuit that starts with a four-qubit entangled state and implements a two-qubit gate on a input two-qubit state ψ by a pair of Bell measurements. See also Ref. (Nielsen, 2003).*

Nielsen's teleportation-based measurement scheme for quantum computation does not rely on an initial entangled state such as a cluster state. All the qubits can be set to a fixed |0⟩ state in the beginning. The Bell states that are needed for teleportation are created by measurement. The measurement needs to involve two qubits simultaneously, unlike the measurements in the one-way quantum computer that only involve individual qubits. In such a teleportation-based scheme, implementation of a one-qubit gate requires two-qubit measurement and that of a two-qubit gate seemingly requires four-qubit measurement. From a different viewpoint, the multi-qubit measurement allows the creation of the needed entanglement. Conceptually, Nielsen's result may be regarded as a simple corollary from the one-way model of Raussendorf and Briegel (Raussendorf & Briegel, 2001). The ability to perform arbitrary 4-qubit measurements means that a cluster state on the honeycomb lattice can be created by measuring its so-called stabilizer operators, which define the cluster state model. The execution of subsequent computation can then be done by one-qubit measurements as in the one-way model.

The requirement of a four-qubit measurement in Nielsen's scheme for the CNOT gate may not be feasible. Fenner and Zhang later reduced the required measurement to three qubits (Fenner & Zhang, 2001), and subsequently, Leung reduced it further to two qubits (Leung, 2001, 2004). Using only two-qubit measurements for universal quantum computation is already optimal in terms of the number of qubits that need to be measured simultaneously.

Later, Aliferis and Leung (Aliferis & Leung, 2004) showed that the teleportation-based approach is equivalent to the one-way approach by demonstrating local mapping between them in the set of universal gates. Subsequently, Childs, Leung, and Nielsen (Childs et al., 2005) used the approach of the one-bit teleportation (Zhou et al., 2000) to unify the two models; see Fig. 5.

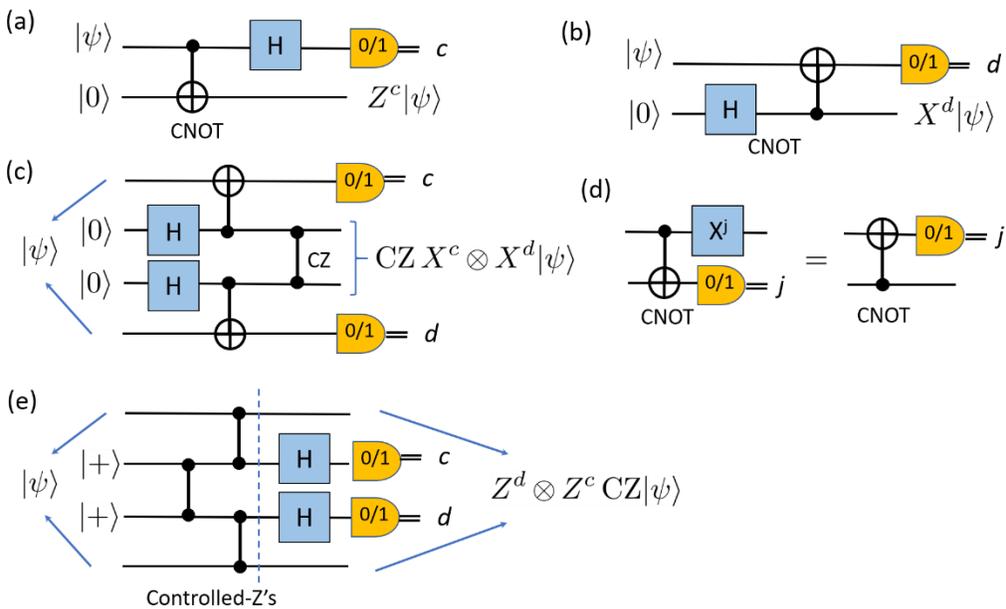

*Figure 5. One-bit teleportation picture. (a) Z-teleportation. (b) X-teleportation. (c) Two X-teleportation followed by a CZ gate. (d) A useful gate identity that swaps the two output ports. (e) Simulation of the Controlled-Z gate. By using circuit identities, including that in (d), it can be shown that (c) is converted to (e). The latter is useful as the action of the CZ gate on a two-qubit input ψ arises from the measurement on the cluster-state entanglement, indicated by the part of the circuit before the vertical dashed line. See also Ref. (Childs et al., 2005).*

*State transfer-based measurement scheme for quantum computation.* Instead of teleportation, Perdrix proposed a state-transfer approach for measurement-based quantum computation (Perdrix, 2005), where only single-qubit and two-qubit observables are used. All observables he used have two outcomes 0 or 1 (or equivalently +1 and -1). The basic state-transfer scheme is shown in Figure 6(a), where each box depicts an observable that represents a two-outcome measurement that projects onto the +1 and -1 subspace of the observable's eigenstates. Unlike teleportation, it uses only two qubits to transfer a one-qubit state.  It was shown that arbitrary one-qubit gates can be implemented by rotating the observables in the state transfer and that the CNOT gate can be implemented by combining two such state transfers with only one auxiliary qubit; see Figure 6(b) and (c). Jorrand and Perdrix used this state-transfer perspective to relate the one-way and teleportation-based approaches in the context of a one-dimensional cluster state (Jorrand & Perdrix, 2005).

Given that universality can be achieved by two-qubit measurements in both the state-transfer picture and the teleportation picture, it seems natural to ask which two-qubit measurements are easier to implement: those of Leung (Leung, 2001) or those of Perdrix (Perdrix, 2005). The answer may depend on physical systems and how the measurements can be implemented.

Beyond the state-transfer picture of computation, it is worth noting that Perdrix and Jorrand also presented a measurement-based approach to construct quantum Turing machines (Perdrix & Jorrand, 2004). The classical Turing machine is a fundamental model of computation that inspires many developments, and its generalization to the quantum regime can also be useful and may lead to further development.

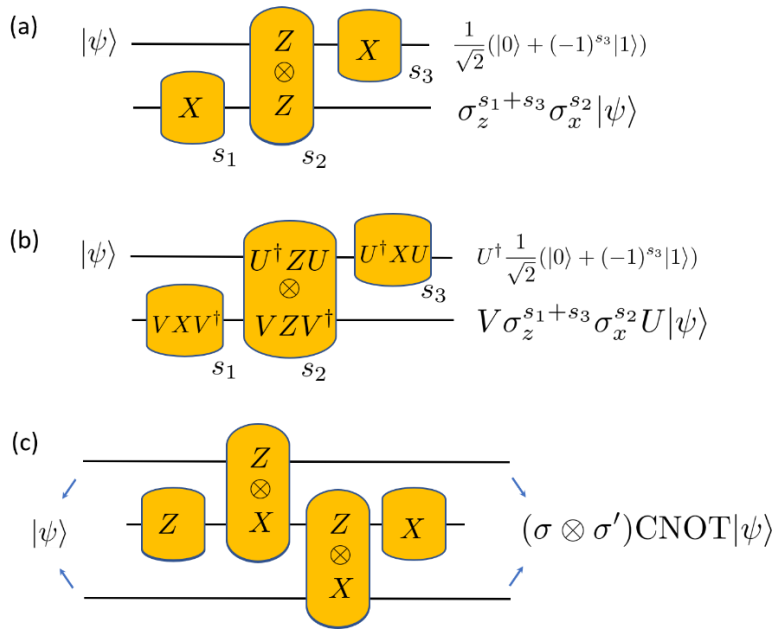

*Figure 6. Measurement-based quantum computation via state transfer. (a) The standard state transfer protocol. Each box represents a two-outcome measurement. For example, the X symbol indicates the measurement that projects onto +1 (s=0) and -1 (s=1) eigenstate of $\sigma_x$, i.e., $(|0\rangle + (-1)^s|1\rangle)|0\rangle/\sqrt{2}$. In terms of operator, each box is a projection operator, such as $(I + (-1)^s \sigma_x)/2$ for the X box, and $(I + (-1)^s \sigma_z \otimes \sigma_z)/2$ for the $Z \otimes Z$ box. By going through a sequence of the three projections, the form of the output shown on the second line can be verified. (b) State transfer-based one-qubit gates. To induce a nontrivial gate on the output, the measurement operators can be transformed by some unitary transformation U and V. (c) State transfer-*

based two-qubit CNOT gate. To implement the action of a CNOT gate, only one additional qubit is needed, with a sequence of four projections (two are single-qubit and the other two are two-qubit). See also Ref. (Perdrix, 2005).

*Valence-bond or correlation-space picture.* Verstraete and Cirac used the picture of valence-bond states (Verstraete & Cirac, 2004b) to understand the one-way computer. The cluster state that Raussendorf and Briegel introduced has an interpretation in terms of a tensor network of valence bonds, or what Verstraete and Criac referred to as projected entangled-pairs states. There are four virtual qubits at each site, except at the boundary, and two neighboring virtual qubits form a maximally entangled pair or a kind of valence bond, $CZ|++\rangle = (|00\rangle + |01\rangle + |10\rangle - |11\rangle)/2 = (|0+\rangle + |1-\rangle)/\sqrt{2}$; see Fig. 7(a). Because each physical site is also a qubit, there is a mapping from the onsite four virtual qubits to one physical qubit via $|0000\rangle \to |0\rangle$ and $|1111\rangle \to |1\rangle$, i.e. a repetition code. A general projected-entangled-pair state can have more general local mapping beyond the repetition code. As depicted in Figure 7(b) and (c), the computation takes place at the virtual qubits and uses teleportation similar but not identical to what was done in (Gottesman & Chuang, 1999; Nielsen, 2003). This approach later instigated the development of the correlation-space MBQC by Gross and Eisert (D Gross & Eisert, 2007). The correlation-space MBQC exploits the tensor-network structure of the states, such as the one-dimensional matrix-product states (Perez-Garcia et al., 2007, p.) as well as the two-dimensional projected-entangled-pair states (Verstraete & Cirac, 2004a, 2004b). It should be pointed out that projected-entangled-pair states and tensor-network states are used almost synonymously in the literature.

For example, Affleck, Kennedy, Lieb and Tasaki (AKLT) constructed a one-dimensional spin chain (Affleck et al., 1987) whose ground state can be written in terms of the matrix-product states, with local matrices corresponding to "+1", "0", "-1" being $A_{+1} = \sqrt{2}\begin{pmatrix} 0 & 1 \\ 0 & 0 \end{pmatrix}, A_0 = \begin{pmatrix} 1 & 0 \\ 0 & -1 \end{pmatrix}, A_{-1} = -\sqrt{2}\begin{pmatrix} 0 & 0 \\ 1 & 0 \end{pmatrix}$, respectively. These matrices represent the respective action on the virtual qubits when a physical spin is measured in the "+1", "0", and "-1" basis. The quantum state of the whole chain can be expressed in terms of the *matrix-product representation*: $|\psi_{AKLT}\rangle = \sum_s \text{Tr}(A_{s_1} A_{s_2} \dots A_{s_N}) |s_2, s_2 \cdots s_N\rangle$. More sophisticated gate actions can be obtained by measuring the physical spin in a general basis; for example, if the measurement projects the physical spin to $(|+1\rangle - |-1\rangle)/\sqrt{2}$, then the gate is proportional to $A_{+1} - A_{-1} = \begin{pmatrix} 0 & 1 \\ 1 & 0 \end{pmatrix} = \sigma_x$, a NOT gate. Extending this example to arbitrary measurement axes leads to a general set of gates that can be implemented by measuring this AKLT state locally. Two dimensions are more complicated, but careful analysis on interesting known states or modification of their local tensors leads to useful gate constructions (David Gross et al., 2007).

After the discussions of the original one-way computer and other variants of measurement-based quantum computation, it is appropriate to point out that in the literature, measurement-based quantum computation, one-way quantum computation, and cluster-state quantum computation are often used synonymously. The subtle difference may lie in what resource states are used and whether measurement is performed on individual qubits or multiple qubits jointly.

**Entanglement in the circuit model and the measurement-based models**

In measurement-based quantum computation, entanglement is the essential resource. In the circuit model, large entanglement may be created during the computation. However, it should be mentioned that in the circuit model, it is possible to realize universal quantum computation with little entanglement, as shown by Van den Nest (Maarten Van den Nest, 2013). The idea is that any circuit $U$ that performs a computation can always be modified by appending an ancillary qubit that is initialized in a state $\sqrt{1-\varepsilon}|0\rangle + \sqrt{\varepsilon}|1\rangle$ and the original action of $U$ is applied when this ancillary qubit is in the state $|1\rangle$. Thus,

the state of the whole system, after such a controlled action, becomes a superposition of (a) the ancillary qubit being $|0\rangle$ and no computation being executed and (b) the ancillary qubit being $|1\rangle$ and the computation $U$ being executed. Because $\varepsilon$ is small, the state of the quantum computer is dominated by case (a) and has little entanglement at any stage of computation. In contrast, the one-way quantum computer requires the substantial presence of initial entanglement in the resource state.

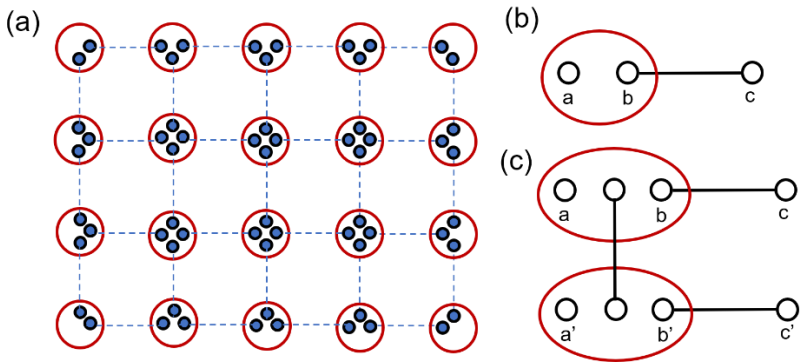

*Figure 7. Tensor network and gates in the cluster state. (a) A valence-bond or tensor-network view of the cluster state. Each site contains a number of `virtual' qubits (equal to the number of neighbors), and each virtual qubit is entangled in the form of $CZ\,|++\rangle = (|00\rangle + |01\rangle + |10\rangle - |11\rangle)/2 = (|0+\rangle + |1-\rangle)/\sqrt{2}$ with one other virtual qubit on a neighboring site. This form of entanglement, as well as the singlet $(|01\rangle - |10\rangle)/\sqrt{2}$ is referred to as a valence bond. The physical qubit is obtained by projecting on to the two-dimensional subspace of $|0...0\rangle$ and $|1...1\rangle$, as if the physical qubit is encoded by a few onsite virtual qubits using the repetition code. Therefore, such a state is also called a **projected-entangled-pair state**. The connected array of virtual qubits with onsite projection can be described by a tensor network, and hence it is also called a tensor-network state. (b) One-qubit gate in the virtual qubit via teleportation. This is a setup similar to quantum teleportation shown in Fig. 4(a). The state on the virtual qubit a can be teleported to c with an additional unitary action. However, the measurement can only be done on the physical qubit, and hence may not be in arbitrary rotated Bell basis of the two virtual qubits. (c) Two-qubit gate via teleportation of virtual qubits. This is similar to a pair of teleportations, except that there is an additional valence bond between the two groups of virtual qubits in the ovals. Via two teleportations, a two-qubit gate can be implemented, which is similar, though not identical, to the setup in Fig. 4(c). See also Ref. (Verstraete & Cirac, 2004b).*

## Part two: Further developments of MBQC and connections to other subjects
### Resource states beyond cluster states

Cluster states are recognized as a resource for measurement-based quantum computation, in particular, in the one-way quantum computer and the correlation-space approach. This was originally shown for the square-lattice cluster state. In fact, cluster states can be defined on any graph, usually referred to as graph states. An immediate question after the work of Raussendorf and Briegel was whether cluster states defined on other 2D lattices were also universal in the sense that they could also be used for universal quantum computation by measuring individual spins. This was first addressed by Van den Nest and collaborators (Maarten Van den Nest et al., 2006), who showed that cluster states on other regular lattices such as the triangular, hexagonal, and kagome lattices, are also universal. This can be intuitively understood by the picture of measurement "LEGO" pieces for universal gates (discussed earlier). Another approach to proving the universality is to demonstrate that these cluster states can be interconverted (to a smaller size) by performing single-qubit measurements on a subset of qubits, as done by Van den Nest and collaborators.

A natural next question is whether the universality holds when the lattice is not perfectly regular or, more generally, the qubits reside on vertices of planar random graphs. Browne and collaborators first addressed this, showing that the universality of the faulty square-lattice cluster state depends on the connectivity of the lattice, or more explicitly, the so-called site percolation threshold (Browne et al., 2008). Such a view of percolation was later shown to hold generally for graph states on planar random graphs (Wei et al., 2012).

Several obvious questions arise. Are there other types of resource states? Can these resource states emerge as ground states of short-ranged Hamiltonians, preferably with a gap? Can thermal states provide useful computation? What is the entanglement requirement of resource states? Can MBQC be fault-tolerant, just like the circuit model employing quantum error-correction codes? Can universal quantum computation become a property of a phase of matter? Is MBQC a practical approach to build a quantum computer? The second part of this review discusses answers to these questions as well as other topics of MBQC.

**MBQC is programmable**.

Nielsen and Chuang showed that it is not possible to build a general-purpose quantum computer to perform an arbitrary quantum computation unless the gate array is operated probabilistically (Nielsen & Chuang, 1997). Their result is based on the circuit model and teleportation. The framework of the MBQC actually allows for a general-purpose quantum computer. In terms of the cluster state, the gate can be applied deterministically provided feedforward is permitted and the size of the cluster state is sufficiently large. The resultant quantum state before the final readout is correct up to Pauli corrections, but the classical outcomes can be corrected. Therefore, it can be argued that such a general-purpose measurement-based quantum computation does allow for arbitrary quantum computation and is hence programmable.

**Entanglement requirement of one-way and correlation-space MBQC**

Systems of limited entanglement can be efficiently simulated by classical computers (Vidal, 2003). From this perspective, entanglement in the universal resource states should grow with their system size, as shown by Van den Nest and coworkers, and is consistent with the entanglement in various universal cluster states (Maarten Van den Nest et al., 2006). Van den Nest and coworkers further applied an entanglement quantifier called Schmidt rank, which is the least number of components in a product form (with respect to a bi-partitioning A:B) that a quantum state can be decomposed to, i.e., the number $x$ in the decomposition $|\Psi_{AB}\rangle = \Sigma_{i=1}^{x} |\psi_i\rangle_A \otimes |\varphi_i\rangle_B$. They showed that when the Schmidt rank of a quantum state, maximized over all bi-partitions, is only logarithmic in the system size, then the efficient classical simulations of MBQC using the quantum state is possible (M. Van den Nest, Dür, Vidal, et al., 2007). This is a no-go result for universal quantum computation with limited entanglement. Thus, it is natural to ask how much entanglement in the resource state is needed for universal MBQC. It is expected to scale with the number of qubits. However, the following result is unexpected.

*Too much entanglement is useless*. Gross, Flammia, and Eisert (David Gross et al., 2009) found that random states generically have a high amount of entanglement and if the entanglement of a quantum state is too high, then using it for MBQC cannot offer any speedup for computation and is no better than random coin tossing. A similar conclusion that random states drawn uniformly from the state space (or in a more technical term, from the Haar measure) are useless for MBQC was reached by Bremner, Mora, and Winter (Bremner et al., 2009). Both results suggest that quantum states that are a universal resource for QC are actually rare and that as commented by Bacon, "entanglement, like most good things in life, must be consumed in moderation" (Bacon, 2009). In fact, by using computational complexity theory, Morimae showed that it is generically a difficult problem to find resource states for measurement-based quantum computation (Morimae, 2017).

**Fault Tolerance of MBQC**

In order to guarantee that quantum computation can proceed as long as is needed, error correction and fault tolerance are necessary. In the circuit model, transversal error correction codes are used to encode a logical qubit by several physical qubits, so that an error can be suppressed at the encoded logical level if the error rate at the physical level is sufficiently low (Gottesman, 1997; Lidar & Brun, 2013). Error correction in other models of quantum computation, such as the adiabatic quantum computation, is still not yet settled. The issue of fault tolerance in the one-way quantum computer was first addressed by Raussendorf in his PhD thesis (Raussendorf, 2003). One can essentially use the 2D cluster state to simulate 1D fault-tolerant circuits. In a similar way, Nielsen proposed to use the teleportation-based approach to simulate quantum circuits with error correction. He argued that a similar threshold theorem should hold here.

Later, Nielsen and Dawson addressed the issue of fault-tolerance in the one-way quantum computation with cluster states (Nielsen & Dawson, 2005). They employed the techniques in the conventional circuit model and developed methods to translate the noise and error considerations into the one-way quantum-computer model. They proved that it is indeed possible that the computation is fault-tolerant, provided the error rate is below a certain threshold. However, they did not give a numerical estimate of the threshold value.

Raussendorf, Harrington, and Goyal (Raussendorf et al., 2006, 2007) exploited a three-dimensional cluster state so that each two-dimensional slice is used to simulate the surface code, a popular error-correcting code (S. B. Bravyi & Kitaev, 1998; Fowler et al., 2009; Kitaev, 2003). However, the surface code alone cannot achieve all universal gates; additional gates that are needed to complete the universality can be inserted by the so-called magic-state distillation (S. Bravyi & Kitaev, 2005). The 3d cluster state can be imagined to be measured layer by layer. Specific measurement patterns mimic the braiding of anyons of topological quantum computation to create gates allowed in the surface code, and others are used to inject the magic state. They showed that the error threshold in this topologically simulated fashion achieved as high as 0.75%, compared to other estimates of order 0.01% or lower (Nielsen & Chuang, 2002). The higher the threshold, the higher the tolerance of errors. Such a topological protection of the MBQC also gives rise to a high threshold in the so-called surface-code quantum computation (Fowler et al., 2012; Raussendorf & Harrington, 2007), intensively pursued in the circuit-model-based quantum computers using a two-dimensional architecture.

Recently, Brown and Roberts developed a general framework that translates a fault-tolerant procedure for stabilizer codes to a measurement-based protocol (Brown & Roberts, 2020) by treating the resource state and single-qubit measurement pattern in the MBQC as a gauge fixing, which is an advanced technique in the subsystem error-correction codes.

**Resource states as ground states of short-ranged interacting Hamiltonians**

Cluster states can be created by unitary evolution induced by Ising-type spin-spin interaction. This was demonstrated in cold atoms (Mandel et al., 2003). However, it may not be easy to achieve such active coupling for other types of resource states. An alternative method, if the resource state is the unique ground state of a short-ranged interacting Hamiltonian with a finite spectral gap, is by cooling the system to low-enough temperature. Unfortunately, cluster states are not unique ground states of any two-body interacting Hamiltonians (Nielsen, 2006). The cluster state on the square lattice is the unique ground state of a five-body interacting Hamiltonian with a nonzero spectral gap. Interaction involving more than two spins is generally difficult to engineer. [If the condition of being exact ground states is relaxed, then the cluster state in certain encoding forms can be an approximate ground state of a two-body interacting Hamiltonian (Bartlett & Rudolph, 2006).] A linear-optical simulation of the cooling of a cluster-state Hamiltonian has actually been performed for a three-site chain, whose Hamiltonian involves only the three spins. Ideally the range of interaction should involve just the nearest neighbors. If such a Hamiltonian can be engineered (which, in itself, is also not a trivial task), then simply 'cooling' the system to low-enough temperature can prepare the system to be close to the perfect universal resource ground state. An obvious question is where such states and their Hamiltonian can be found.

The first provable universal resource state with a nearest-neighbor interacting parent Hamiltonian with a non-zero spectral gap is the so-called tri-cluster state defined on the hexagonal lattice, invented by Chen and collaborators (X. Chen et al., 2009). This is a quantum state with a local Hilbert space of dimension six, which contains the cluster state in three different bases, hence the name tri-cluster state. Despite this having more than two levels, the tri-cluster state can be further converted to a cluster state of qubit local Hilbert space (i.e. of two levels) by the so-called quantum state reduction (X. Chen et al., 2010).

**Tensor-network states and correlation-space MBQC**

The correlation-space measurement-based quantum computation taps into tensor-network states for the enabling resource (D Gross & Eisert, 2007; David Gross et al., 2007; Verstraete & Cirac, 2004b) (D Gross & Eisert, 2007; David Gross et al., 2007; Verstraete & Cirac, 2004). It explains how the cluster state used in the one-way quantum computer can be understood with local tensors. It offers a simple explanation of local gates and also generalizes resource states by modifying local tensors. However, it should be pointed out that the computation is carried out in the Hilbert space of virtual qubits, in contrast to the one-way quantum computer where the computation is done in the Hilbert space of physical qubits. Some example states investigated in the correlation-space picture include the AKLT state and modified toric code states (David Gross et al., 2007).

**Affleck-Kennedy-Lieb-Tasaki states for universal MBQC**.

One family of states that has gained much attention for the MBQC is the one constructed by Affleck, Kennedy, Lieb, and Tasaki (AKLT) (Affleck et al., 1987, 1988). The particular 1D AKLT model gives strong evidence of Haldane's conjecture (Haldane, 1983) that isotropic quantum spin chains of integer spin have a unique ground state with a finite spectral gap. This is the opposite of half-integer spin chains, where the ground state is either degenerate or the system does not possess a finite spectral gap (Lieb et al., 1961). The AKLT construction by valence-bond states naturally generalizes to higher dimensions and arbitrary graphs. It was shown that these AKLT states are unique ground states of certain isotropic two-body interacting Hamiltonians. The local Hilbert-space dimension and the explicit form of the Hamiltonian depend on the local structure of a lattice.

The 1D AKLT state of local Hilbert-space dimension 3 (i.e. qutrits) was first explored by Gross and Eisert in the measurement-based quantum computation (D Gross & Eisert, 2007; David Gross et al., 2007) using the correlation-space picture. Brennen and Miyake (Brennen & Miyake, 2008) later realized that, to execute one-qubit operation in the edge state of the spin-1 AKLT chain, the coupling of the edge spin with the bulk must be turned off and a subsequent local measurement performed on it. In fact, this works with any spin chain in the so-called Haldane's phase that is symmetry protected (Miyake, 2010).

To go beyond one dimension, Cai and coworkers considered stacked layers of 1D AKLT chains with decoration; namely in each layer there are spins of local dimension 4 residing on the backbone of a chain and spins of local dimension 2 are connected to each site of the backbone. They transformed such a layer structure of 1D chains into a 2D AKLT-like state. They showed that this state is universal for MBQC (Cai et al., 2010). Later it was shown by Wei and coworkers (Wei et al., 2011) and independently by Miyake (Miyake, 2011) that the original 2D spin-3/2 AKLT state on the hexagonal lattice is actually universal for MBQC. Such a result was also generalized beyond the hexagonal lattice (Wei, 2013; Wei, Haghnegahdar, et al., 2014), including the universality of the spin-2 AKLT state on the square lattice (Wei & Raussendorf, 2015).

One approach to show that AKLT states are universal for MBQC is to convert the AKLT state to a cluster state, which is itself universal, via local measurement. In the case of the spin-3/2 AKLT states, a four-level system must be mapped locally to a two-level system. This can be achieved by a generalized measurement at all sites. Similar to the projective measurement, the outcome of the generalized measurement on the AKLT spins is also random and has three different outcomes labeled by x, y, or z. It was shown that for any outcome of the generalized measurement on all sites, the AKLT state is transformed into an encoded graph state. Encoding simply means that a logical qubit is extended to connected sites of the same type of outcome (x, y, or z); see Fig. 8(b)(e). The graph is modified from the hexagonal lattice: each domain that contains connected sites of the same outcome form a vertex, whereas the interdomain edges need to be treated in a modulo-2 manner: an even number of edges will be converted to no edge between two domains, but an odd number of edges will be converted to a single edge that connects two domains; see Fig. 8. Invoking the results of universality for random planar graphs, if their connectivity as defined by percolation is sufficiently high, then the graph states are as good as regular cluster states for MBQC. This connectivity was checked and confirmed by numerical percolation simulations (Wei et al., 2011).

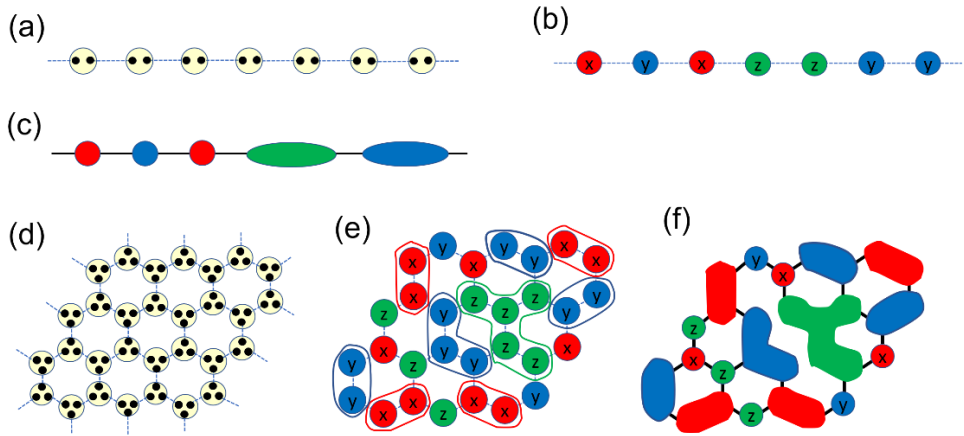

*Figure 8. The preprocessing generalized measurement on the one-dimensional and hexagonal AKLT states. (a) & (d): valence-bond definition of the AKLT states. Each site consists of 2 or 3 virtual qubits, depending on the number of neighbors, and two neighboring qubits form a valence-bond state of the form $(|01\rangle - |10\rangle)/\sqrt{2}$. A physical spin is obtained from the virtual qubits by symmetrization, e.g. 00→ "+1", 01+10→ "0", and 11→"-1", where "+1", "0", and "-1" are the labels for the physical spin on a site of a linear chain. (b)&(e) illustrate the random outcomes of the generalized measurement on all sites; there are three possible outcomes labeled by X, Y and Z. For example, in (b) the Z indicates that the measurement projects the local site to a two-dimensional Hilbert space spanned by "+1" and "-1", and the X and Y indicate similar projections but rotated from Z axis to X and Y axes, respectively. Similar generalized measurement is also performed on the hexagonal lattice, indicated in (e). (f) An example of domains, which contain connected sites with the same outcome of the generalized measurement. (c) & (f) are the resultant graphs for the graph states to which the AKLT states are converted by the generalized measurement. As seen in (c) & (f), some blocks (or also known as domains) are composed of a few sites, due to the valence-bond correlation that gives rise to a redundant encoding of a logical qubit by a few physical sites, when these connected sites share the same outcome. The generalized measurement filters out a graph state randomly from the AKLT state.*

Another approach to proving the universality is to demonstrate that universal gates can be simulated. Miyake used the same generalized measurement and defined the notion of a computational backbone (Miyake, 2011), where one- and two-qubit gates were constructed in the correlation-space picture. He argued that a macroscopic size of the backbone exists with a sufficiently high probability on the hexagonal lattice, and thus, the AKLT state is universal for MBQC.

Higher spins present specific technical difficulties. However, Wei and Raussendorf managed to show that the spin-2 AKLT state on the square lattice is universal (Wei & Raussendorf, 2015, p. 2). Whether AKLT states with higher spins than 2 are universal for MBQC remains open.

The issue of the nonzero gap above the ground state in the spin-3/2 model on the hexagonal lattice has been a longstanding question. AKLT showed that the spatial correlation function in the ground state decays exponentially, but the existence of the gap could not be proved (Affleck et al., 1987). Recently, two groups independently used numerically assisted approaches to show that the AKLT model indeed possesses a nonzero spectral gap (Lemm et al., 2020; Pomata & Wei, 2020), even in the limit that the system size becomes infinite. Therefore, the AKLT models provide example Hamiltonians that are short-ranged, gapped, and have a unique ground state that is universal for measurement-based quantum computation. This property may be helpful when creation of the ground-resource states is performed by cooling the temperature of the physical system.

**Symmetry-protected topological states and quantum computational phases of matter**.

The lack of a systematic approach to characterize universal resource states has led researchers to consider certain phases of matter, and in particular, the symmetry-protected topological phases. Else and coworkers (Else et al., 2012) found that teleportation of the one-qubit state is possible in the correlation space anywhere within a symmetric phase of $Z_2 \times Z_2$, but general gates can only be achieved at very special points in the phase of matter. The $Z_2$ symmetry group consists of only two elements, such as the identity element and a rotation around x axis by 180 degrees; $Z_2 \times Z_2$ is a symmetry group that is a product of two such $Z_2$ symmetry groups (that commute with each other). Example states in the nontrivial $Z_2 \times Z_2$ phase include the 1D cluster state and the 1D AKLT state. The ability to implement teleportation in a quantum wire with $Z_2 \times Z_2$ symmetry (as in the work of Else et al.) was later extended to other symmetry groups, including non-Abelian ones (Prakash & Wei, 2015). More relevantly, Miller and Miyake generalized the idea of renormalization (Bartlett et al., 2010) and used it to show that the 1D symmetry-protected topological phase by $S_4$ symmetry (which is the permutation group of 4 objects) can give rise to the implementation of arbitrary one-qubit gates (Miller & Miyake, 2015). Subsequently, Stephen and coworkers extended this more generally (Stephen et al., 2017). This is the strongest connection of symmetry-protected topological phases to quantum computation. However, a one-dimensional state of matter only offers limited computation, such as one-qubit gates. In order to obtain universal quantum computation, higher dimensions are needed.

Doherty and Bartlett considered teleportation to be a necessary condition and devised an order parameter to detect it in a cluster Hamiltonian with an external field (Doherty & Bartlett, 2009). They found that such characterization coincided with the conventional phase diagram of the model. However, the ability to teleport does not necessarily imply the ability to implement universal gates.

Going beyond one dimension, Poulsen-Nautrup and Wei considered the fixed-point wavefunctions of 2D symmetry-protected topological phases constructed by Chen and coworkers using the mathematics of cohomology and showed that they could be used to perform universal measurement-based quantum computation (Poulsen Nautrup & Wei, 2015). Independently, Miller and Miyake considered a different symmetry-protected topological state (with $Z_2 \times Z_2 \times Z_2$ symmetry) on the "union-jack" lattice based on a Control-Control-Z gate construction by Yoshida (Yoshida, 2016) and showed that this state could also be used for universal measurement-based quantum computation (Miller & Miyake, 2016). This universality was later generalized to the symmetry of $Z_d \times Z_d \times Z_d$ (Y. Chen et al., 2017). One interesting feature in the work of Miller and Miyake is that universality can already be achieved by measuring Pauli operators, namely along the x-, y- and z-axis of the Bloch sphere, which is not the case in the cluster state. in Ref. (Wei, 2018), the construction by Miller and Miyake was shown to be equivalent to a different, but widely known, topological state constructed by Levin and Gu (Levin & Gu, 2012), whose model was a paradigmatic one for two-dimensional symmetry-protected topological phases. However, these studies only apply to specific representative wavefunctions of the symmetry-projected topological phases. An attempt was made by Wei and Huang that extended the universality to an extended region around some of these fixed-point states (Wei & Huang, 2017), but whether an entire phase could be reached was not known at that time.

It is possible to obtain universal resource from an entire phase of matter in two dimensions. The particular phase is called the cluster phase (Raussendorf et al., 2019), which contains the cluster state as a specific example. It has been studied on various 2D lattices (Daniel et al., 2020; Devakul & Williamson, 2018), and it was understood that the essential symmetry that provides such computational power belongs to the so-called subsystem symmetry, including a symmetry element which acts on spins located spatially in a fractal pattern. These results point to a possible general notion of quantum-computational phases of matter. In fact, a different perspective of quantum-computational phases of matter has been explored in the context of intrinsic topological phases where braiding of anyonic excitations leads to a myriad of quantum gates (Nayak et al., 2008).

**Thermal states for measurement-based quantum computation**.

The cluster state in the one-way quantum computer can be regarded as the ground state of a cluster Hamiltonian, which is related to a simple paramagnetic Hamiltonian via transformation using Controlled-Z gates (Briegel & Raussendorf, 2001). The ground state is the property of a system at zero temperature, but in real life, the system will always sit at a finite temperature. Thus, it is natural to consider one-way computation at finite temperatures. Fujii and coworkers compared the cluster Hamiltonian and a related interacting cluster Hamiltonian that is transformed to an Ising-interacting Hamiltonian and investigated the finite-temperature effect on the computational capability (Fujii et al., 2013). The latter model possesses a thermal phase transition, whereas there is no transition in the original cluster-state model. Fujii et al. found that the long-range order in their model enhances the robustness of quantum computation against thermal excitations. In going beyond cluster models, Li and coworkers constructed two models in two- and three-dimensions in which the thermal states are useful for universal MBQC and the interactions do not need to be turned off during computation (Li et al., 2011). The three-dimensional model was subsequently modified by Fujii and Morimae to one that possesses uniform spin-3/2 entities on all sites. They showed that from the thermal state, a relatively clean cluster state of high connectivity could be distilled (Fujii & Morimae, 2012). Other constructions were proposed (Wei, Li, et al., 2014) that also discussed the thermal transition of quantum-computational power. Consideration of thermal states and the finite-temperature effect for measurement-based quantum computation will become relevant in the effort of building a realistic measurement-based quantum computer.

**MBQC and classical computation**

The aim of measurement-based quantum computation is to achieve the capability of universal quantum computation. It not only relies on simple classical computation but may also yield insight on the latter. Van den Nest and Briegel established a connection between the MBQC and the field of mathematical logic (Maarten Van den Nest & Briegel, 2008). In particular, if a graph state yields a speed-up of the quantum computation with respect to its classical counterpart, then the underlying graph is associated with an undecidable logic theory, where the undecidability is similar to Gödel's incompleteness results.

From a different perspective, Anders and Browne studied how the correlations exploited in MBQC enabled computational power (Anders & Browne, 2009); see Fig. 9. Cluster states possess certain kinds of entanglement and correlations, and the classical computer interacting with such correlations (as revealed by measurement) only needs to execute binary addition in order to achieve universal quantum computation. Thus, a meaningful question is, with the limited power of a classical computer, how do the correlations give rise to computational power? For certain tensor-network states, to achieve universal quantum computation, the classical computer needs operations beyond binary addition. Said conversely, a limitation to perform only binary addition (i.e. parity) for the classical computer interacting with the correlations from the tensor network states may not achieve universal quantum computation. This also leads to the concept of measurement-based classical computation: what kind of correlations can boost the computational power of a classical parity computer? Anders and Browne showed that correlations in any bipartite quantum states cannot help to realize the classical NAND gate deterministically. In contrast, the three-qubit GHZ can do that, thereby boosting the classical computer to a classical universal one. These considerations also reveal a connection between the violation of local realistic models and the computational power of entangled states. Such violation is a manifestation of the so-called *contextuality* in the foundations of quantum mechanics (Kochen & Specker, 1967). Naively, one might expect that the measurement of observables simply reveals their pre-existing values and hence is not *contextual*. However, this view is at odds with quantum mechanics.

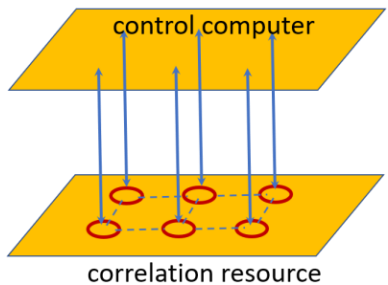

*Figure 9. A control computer interacts with a correlation resource; see also Ref. (Anders & Browne, 2009). This schematic diagram explicitly specifies the classical control computer, which is needed in the measurement-based quantum computation to compute the basis adaptation, such as that indicated in Fig. 3(c). The computational power of the classical control itself is limited. However, it can send the instruction of measurement axis to the correlation resource, which is an entangled state, and reads out the two-outcome measurement 0/1. Depending on the correlation resource, the resultant computational power of the classical computer can be enhanced.*

In addition to its role in quantum foundations, contextuality has been shown to supply the 'magic' to quantum computation (Bermejo-Vega et al., 2017; Howard et al., 2014). It is known that quantum computation with a limited gate set such as the Pauli gates, Hadamard, phase and CNOT gates (in the family of Clifford gates) can be efficiently simulated by a classical computer. A non-Clifford gate is needed to boost the power of a quantum computer. The consequence of a state being contextual is that a magic state can be distilled out of it and enables implementation of non-Clifford gates, making a quantum computer universal. Clifford gates are those that transform a product of Pauli operators to another product form, and quantum computation using only Clifford gates can be efficiently simulated by a classical computer, therefore such a computer cannot achieve universal quantum computation (Gottesman, 1999). An example of a non-Clifford gate is a rotation around the z-axis by 45°, also known as the T gate. Adding this T gate to the set of Clifford gates unleashes the power of universal quantum computation.

Given that contextuality is intimately related to measurement, Raussendorf expanded the study of contextuality in MBQC and showed that such a qubit quantum-computational model with classical binary-addition capability is contextual if it can compute a nonlinear Boolean function with a high probability. Namely, such a computational model cannot be explained by a realistic local hidden-variable model. In particular, this shows that such MBQC executing the quantum algorithm for the discrete log problem is contextual; the super-polynomial speedup over the best-known classical algorithm seems to be supplied by contextuality. Such a result was recently generalized to the qudit (with $d$ levels instead of two) scenario that shows strong non-locality is necessary for MBQC evaluating high-degree polynomial functions, with the classical control computer having only linear processing capability (Frembs et al., 2018).

**Time ordering in MBQC**

In the one-way quantum computer of Raussendorf and Briegel (Raussendorf & Briegel, 2001), measurement axes of some qubits may depend on the measurement outcomes of previously measured qubits. This results in partial time ordering among qubits in terms of measurement (Raussendorf et al., 2003). This can also be formulated in terms of the flow of quantum information (Danos & Kashefi, 2006), as illustrated in Fig. 10, which has led to a flow condition that gives rise to deterministic computation on graph states (de Beaudrap, 2008b, 2008a). Measurement calculus has also been developed for the one-way quantum computer (Danos et al., 2007). These have led to the reduction and parallelization of a certain class of polynomial-depth circuits to logarithmic ones (Broadbent & Kashefi, 2009). The notion of flow has also been generalized so as to deal with the situation where there is no flow on an entanglement graph, but instead a generalized flow exists, as well as to optimize implementation of the unitary gates (Browne et al., 2007). Generalizing this to stabilizer states beyond graph states, temporal relations and measurement settings were classified in terms of bases of the so-called check matrix that characterizes these states. This also gave rise to the result that classical processing relations for deterministic computation can constrain the resource state and measurement setting (Raussendorf et al., 2016).

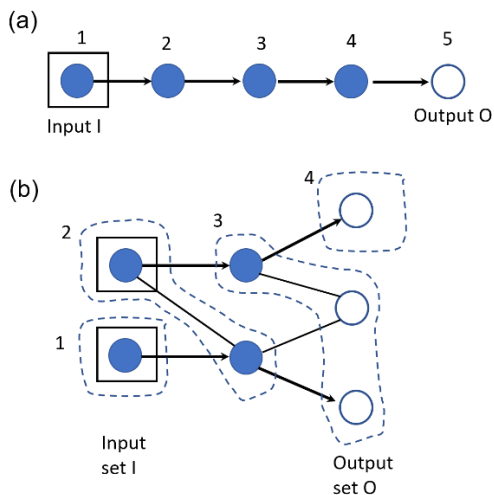

*Figure 10. Illustration of a flow. Dependence is indicated by arrows. A partial order is marked by a number on a group of vertices; see also Ref. (Danos & Kashefi, 2006). There are an input set I of vertices and an output set O of vertices. All qubits, except those in O, will be measured. The complement of I is the set of all vertices not in I and it is denoted by $I^c$, and similarly the complement of O is denoted by $O^c$. A flow consists of (i) a mapping f from $O^c$ to $I^c$, marked by an arrow between neighboring vertices, and (ii) a partial ordering >, so that f(i)>i. In order for the two conditions to be consistent, any neighbor, e.g., k of f(i) that is not i, must be k>i. The existence of a flow ensures that a deterministic unitary gate can be implemented. (a) A one-dimensional graph with a flow. Here the partial order labeling coincides with the qubit labeling. (b) A graph with a flow. The numbers outside the dashed boxes indicate the ordering.*

## MBQC and classical spin models

In statistical mechanics, knowledge of the partition function of a system gives rise to its equilibrium properties (Baxter, 2016; McCoy, 2010). Van den Nest, Dür and Briegel found that the partition function of the well-known classical Ising model in statistical mechanics (Baxter, 2016; McCoy, 2010) can be written as the overlap between a resource state Ψ and a product state (M. Van den Nest, Dür, & Briegel, 2007). The resource state Ψ is a graph state that encodes the interaction pattern of the model, and the product state encodes coupling and local field strengths, which can be complex in general. Such an overlap represents a branch in the measurement-based quantum computation. If it is easy to compute the corresponding partition functions for all model parameters, then the quantum computation can be efficiently simulated by classical means, and thus the corresponding resource state is not universal (Van den Nest et al., 2007). Moreover, the 2D Ising model is regarded as complete in that the partition function of the q-state Potts models in statistical mechanics (Baxter, 2016; McCoy, 2010) can be reduced to an instance of the partition function of the Ising model with generally complex parameters. The connection to MBQC is made via the branch in the computation (specified by the product state) using a 2D cluster state for both the Ising and q-state Potts models in statistical mechanics (M. Van den Nest et al., 2008). Using measurement-based quantum computation to study classical spin models seems to be an interesting research direction. Considerations along this line of thought have led to the fruitful finding that all the physics of every classical spin model is reproduced by certain "universal models" in their low-energy sector and that the two-dimensional Ising model with fields is universal (De las Cuevas & Cubitt, 2016).

## Blind quantum computation.

In the one-way computer, once the resource state and the measurement patterns are fixed, the specific quantum circuit is determined. Imagine a server that takes the instruction of measurement axes and reports the outcomes to a client that intends to run some quantum computation. Is it possible that the client can instruct the server, but the latter cannot find out what quantum circuit has been executed? Broadbent, Fitzsimons, and Kashefi (Broadbent et al., 2009) devised so-called blind quantum computation using measurement-based quantum computation to achieve this. However, it requires the client to prepare the initial product state of the entire array of qubits in the form $|0\rangle + e^{i\theta}|1\rangle$, where the phase $\theta$ is a multiple of $\pi/4$. Then the client sends all qubits to the server, which then places them on a brickwork lattice (see Fig. 11) and applies the Controlled-Z gates pairwise according to the brickwork structure. Subsequent communication between them is entirely classical. They communicate back and forth via the client informing the measurement axes of a column of qubits to be measured, and the server returns the measurement outcomes. The computation terminates when all qubits have been measured. Broadbent and coworkers showed that by randomly initializing the qubits and randomly flipping the measurement axes, the client could hide the computation from the server. A small-scale experimental demonstration of blind quantum computation has been carried out by Barz and coworkers (Barz et al., 2012). There have been many works following up on the idea of blind quantum computation; see the review by Fitzsimons (Fitzsimons, 2017) and references therein.

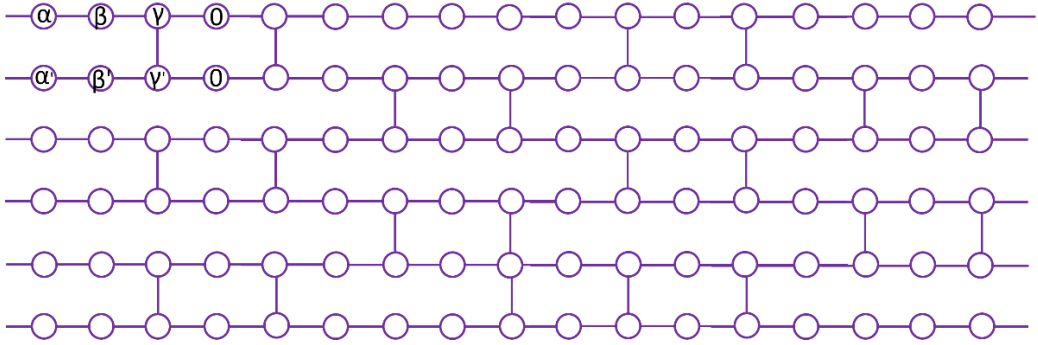

*Figure 11. Brickwork lattice that hosts the blind quantum computation; see also Ref. (Broadbent et al., 2009). Each circle represents a qubit and the symbol inside it indicates the measurement axis or the corresponding observable, e.g., $\hat{O}(\xi) \equiv \cos\xi\,\sigma_x + \sin\xi\,\sigma_y$. The brickwork state is a graph-like state with the graph being the brickwork lattice and is defined by a two-step process: (1) the client prepares each qubit randomly in any of the states: $(|0\rangle + e^{i\theta}|1\rangle)/\sqrt{2}$. with $\theta = 0, \frac{\pi}{4}, \ldots, \frac{7\pi}{4}$, and sends all the qubits to the server; (2) the server arranges all qubits on a brickwork lattice and applies CZ gates pairwise to those two spins connected by an edge. Similar to the cluster state, the computation proceeds by measuring qubits from left to right, with later measurement axis adaptation; in the server-client setting, this is informed by the client to the server that performs the measurement. The achievement of the blind quantum computation is that the client can perform a delegated computation by specifying the measurement axes without the server knowing the computation itself.*

**Issues of measurement axes and observables**

Already in the work of blind quantum computation (Broadbent et al., 2009), measurement of observables in the x-y plane is sufficient, as there is no need to carve the required entanglement structure from some other initial cluster or graph states. Mantri and coworkers consider open-ended rectangular lattices and show that for cluster states on these lattices, measurement in the x-y plane is also sufficient (Mantri et al., 2017). In a $Z_2$ symmetry-protected topological state, which belongs to the hypergraph states, Miller and Miyake showed that only Pauli X, Y, and Z measurements are sufficient (Miller & Miyake, 2016). Subsequently, Takeuchi and coworkers constructed a specific hypergraph state such that only Pauli X and Z measurements are sufficient (Takeuchi et al., 2019). It is believed that further reduction of measurement is unlikely to be possible, but Pauli measurements are relatively easy to implement. However, hypergraph states may not be trivial to generate.

**Linear-optical quantum computation**

The perspective of MBQC also revived the proposal by Knill, Laflamme and Milburn that showed that it is possible to use linear-optical elements assisted with single-photon sources and detectors for universal quantum computation in the standard circuit model (Knill et al., 2001). Despite the scheme being possible in principle, the required resources involved are daunting (Li et al., 2015). It is by using cluster states of the one-way quantum computation (Browne & Rudolph, 2005; Nielsen, 2004) that the interest in linear optical quantum computation was revived, as the resource requirement was dramatically reduced (see Fig. 12). Some small cluster states were realized by merging down-conversion entangled photon pairs (Lu et al., 2007; Walther et al., 2005). There have been further works that propose methods to create 2D cluster states (Economou et al., 2010; Gimeno-Segovia et al., 2019; Lindner & Rudolph, 2009). Recently there has been some experimental effort towards realizing key proposed ingredients (Schwartz et al., 2016).

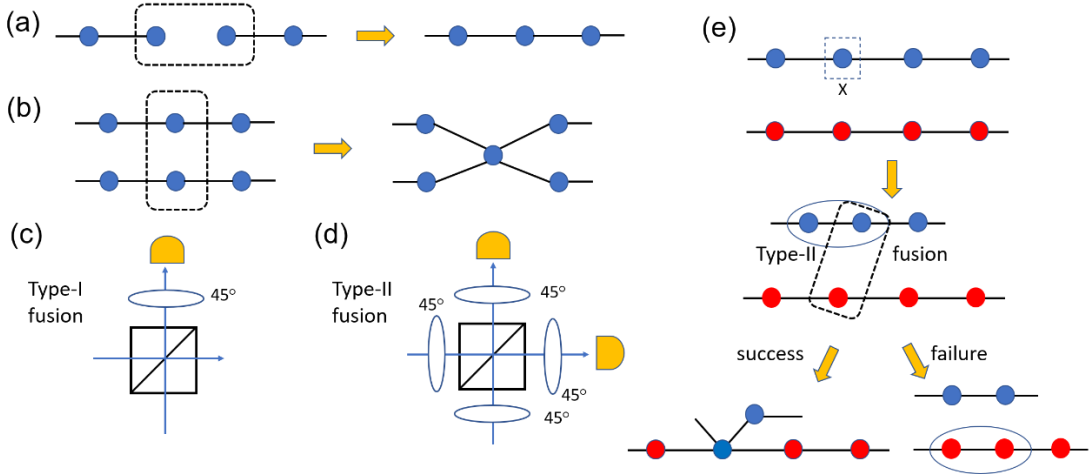

*Figure 12. Schematic illustration of creating a cluster by merging. (a) Merging two pairs into a linear chain. (b) Merging two triplets into a cross. (c) Type-I fusion gate for photons. It consists of a polarizing beam splitter that acts on a pair of photons as well as a 45° polarization rotation before a photon counter. The desired action succeeds with a probability of ½, and when it does, it achieves the merging operations illustrated in (a) and (b). When it fails, it removes the two photons from the clusters and further disconnects each chain into two parts. (d) Type-II fusion gate. Its design consists of four polarization rotations and two photon counters. (e) An application of Type-II fusion. There are two linear clusters. One of the qubits is measured in the X basis (e.g. using a polarizing beam splitter at a 45°), and this joins the two neighboring qubits to form a logical qubit in the repetition code, indicated by the oval. Then a Type-II fusion attempt is made on the two qubits enclosed by the dashed box. When the fusion is successful, it merges the two clusters with reduced sizes. When it fails, it does not break each chain (as would be the case in the Type-I fusion); it simply removes the redundant encoding in the upper chain and joins the two sites into a redundant encoding in the lower chain. See also Ref. (Browne & Rudolph, 2005).*

In addition to using discrete basis states such as polarization or time bins, another related development is to use continuous variables of light, i.e. the continuous degrees of freedom in its electric field. Menicucci and coworkers proposed schemes to generate continuous-variable cluster states (Menicucci, 2014; Menicucci et al., 2007). He later showed that it is possible to use them for fault-tolerant measurement-based quantum computation (Menicucci, 2014). There have been experimental achievements in realizing large-scale cluster states of a large number of optical modes (M. Chen et al., 2014; Larsen et al., 2019; Yokoyama et al., 2013; Yoshikawa et al., 2016). However, it is still a challenge to perform local optical-mode measurement for universal quantum computation.

**Graph states and measurement-based approach for quantum communication**

Bell states can be used to teleport an unknown quantum state, but in order to teleport over a long distance, such a long-distance entanglement needs to be established. If there is an array of Bell pairs distributed across two distant nodes, then so-called entanglement swapping can serve this purpose. As shown in Fig.13(a), with one Bell pair shared between A and B and another shared between B and C, party B performs a Bell-basis measurement and forwards the outcome to C, the initial shared A-B entanglement can be teleported to form entanglement between A and C. This is entanglement swapping (Pan et al., 1998). By applying this to an array of entangled pairs, shown in Fig.13(b), a long-distance entanglement can be established (Sangouard et al., 2011). This is the basic setup of the so-called "quantum repeaters" (Duan et al., 2001). In fact, the measurement-based approach has provided a useful framework to consider ideas from entanglement purification, noisy channels, fault-tolerance, and transmission of big quantum data together (Pirker et al., 2018; Wallnöfer & Dür, 2017; M Zwerger et al., 2012, 2013, 2014, 2016; Michael Zwerger et al., 2018). Some of the proposed methods have been realized experimentally (Chrzanowski et al., 2014), including amplification of degraded entanglement and extraction of secure keys in an otherwise insecure regime.

In using entangled photons there is, however, a limitation due to the finite failure probability of photonic Bell measurement, which is 1/2 without using additional resource (Calsamiglia, 2002). This means that successful long-range entanglement only happens at an exponentially small rate. Azuma, Tamaki and Lo proposed to use cluster states or graph states to solve this issue (Azuma et al., 2015). The graph of the graph states used in this quantum communication scheme consists of inner nodes that form a complete graph and outer nodes (also call leaf nodes) that are connected to the inner nodes. In Fig. 13(c), two such graph states are shown, which replace the two Bell pairs in Fig. 13(a). Because of multiple leaves, multiple attempts of Bell measurement can be made and the success probability that A and B become entangled can be boosted from $1/2$ to $1 - 1/2^n$, where $n$ is the number of leaves. This scheme, in principle, allows quantum communication without using quantum memories to temporarily store the states of photons. However, the challenge is to create such a graph state; one natural approach is to use the fusion schemes in Fig. 12.

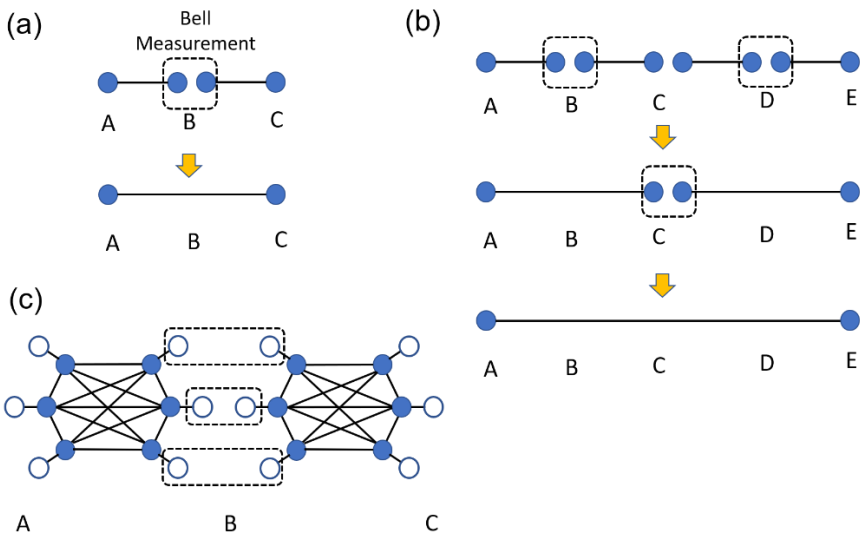

*Figure 13. Entanglement swapping and long-distance entanglement. (a) Basic entanglement swapping uses two pairs of Bell states and Bell measurement. It can be regarded as teleportation of the left qubit of B to the qubit of C and therefore, A and C will share a Bell pair afterwards, despite the fact that A and C were never entangled before. (b) Using (a) as the basic protocol, a long-distance entanglement can be established, e.g. between A and E. However, Bell measurement on photons (without using additional resources) only succeeds half of the time. (c) Generalization of entanglement swapping using more complicated entangled states, such as the graph state. The graph consists of inner nodes that form a complete graph (all nodes are connected pairwise) and outer nodes that are connected to the inner nodes. Two such graph states are shown, and one is shared between A and B, and the other one is shared between B and C. Because of multiple leaves, multiple attempts of Bell measurement can be made and the success probability that A and B become entangled can be boosted from ½ to 1-1/2$^n$, where n is the number of neighboring leaf pairs. The states can be further simplified to a simpler graph by measuring some inner nodes possessed by B.*

**Experimental progress**

Arguably the first experimental realization of a cluster state was done by the group of Bloch using cold atoms trapped in an optical lattice (with two selected hyperfine states as a qubit) (Mandel et al., 2003). They used a 'cold controlled collision' method (Jaksch et al., 1999) already envisaged in the original work of the cluster state by Briegel and Raussendorf (Briegel & Raussendorf, 2001), which shifted atoms by a lattice site depending on their hyperfine spin state so as to induce a phase shift for certain combinations of nearby spin states. However, at that time, individual addressing such as single-atom measurement and gate operation were not possible and implementation of the one-way computer was still very challenging. Recent progress on imaging and addressing of individual atoms makes the realization of the one-way computation in trapped cold atoms probably not far-fetched (Bakr et al., 2009; Edge et al., 2015; Sherson et al., 2010; Simon et al., 2011; Weitenberg et al., 2011). In addition to previous use of bosonic cold atoms, a scheme for cluster-state generation with trapped fermionic atoms using interplay of the spin-orbit coupling and superexchange interaction has also been proposed, which may potentially have longer coherence time (Mamaev et al., 2019).

Instead of the cold collision, a Rydberg state can be exploited to induce a phase shift for two atoms in a particular hyperfine state that is driven resonantly to this Rydberg state. This is due to the interaction of the extended electron clouds of the two atoms in a Rydberg state and is usually referred to as the Rydberg blockade (Jaksch et al., 2000; Lukin et al., 2001; Weiss & Saffman, 2017). Rydberg blockade and entanglement generation between two neutral atoms via the Rydberg blockade have been demonstrated experimentally (Urban et al., 2009; Wilk et al., 2010; Zhang et al., 2010). This has also led to implementation of a Controlled-Z gate and it can potentially be used to directly create a cluster state of an array of atoms (Briegel & Raussendorf, 2001).

Small-size cluster and graph states have also been realized experimentally by probabilistically merging pairs of entangled photons (Lu et al., 2007; Walther et al., 2005); a small graph-state error-correction code was implemented (Bell et al., 2014). Deterministic schemes for their generation have also been proposed using solid-state and quantum-dot emitters (Economou et al., 2010; Gimeno-Segovia et al., 2019; Lindner & Rudolph, 2009). Important ingredients underlying these schemes have also been realized experimentally (Schwartz et al., 2016). In addition to the discrete polarization degrees of freedom of light, the so-called continuous-variable states of light have been employed to create large-scale cluster states in optical modes (M. Chen et al., 2014; Larsen et al., 2019; Yokoyama et al., 2013; Yoshikawa et al., 2016). One challenge for that system to implement computation is the measurement of individual modes and the fast feedforward to adapt subsequent mode measurements.

Cluster and graph states have also been generated in other physical systems, such as in trapped ions, where some error correction codes were created (Lanyon et al., 2013), and in superconducting qubits, where some experiments were performed via the cloud-based publicly available quantum computers of IBM (Mooney et al., 2019; Wang et al., 2018).

Generation of resource states beyond cluster states seems to be harder. Nevertheless, certain one-dimensional tensor-network states used in the correlation-space approach have also been realized (Gao et al., 2011), including a short chain of the AKLT state (Kaltenbaek et al., 2010).

There are other theoretical proposals to produce cluster states and implement measurement-based quantum computation on various physical systems (Cho & Lee, 2005; Guo et al., 2007; Koch-Janusz et al., 2015; Kuznetsova et al., 2012; Lim et al., 2005, 2006; Lin et al., 2008; Tanamoto et al., 2006, 2009; Weinstein et al., 2005). It may be possible that the measurement-based approach will result in practical quantum computers in the not-so-distant future, comparable to those based on the standard circuit model.

## Conclusion

Measurement-based quantum computation offers both an intellectual framework for quantum information processing and a blueprint for potentially building up a quantum computer. For example, the entanglement requirement for computation was explored, and partial time ordering and symmetry were also studied for deterministic computation. Furthermore, how correlations could be used as a resource for classical computation also links to the foundations of quantum mechanics. Universal blind quantum computation was an unexpected application of measurement-based quantum computation, which could be useful in future secure cloud-based quantum computation. In fact, application of the measurement-based approach to quantum communication is already feasible. From the perspective of condensed matter, the existence of an entire phase of matter capable of universal quantum computation makes the notion of the quantum-computational phase of matter an interesting new interdisciplinary direction to explore. The establishment of fault tolerance in the MBQC and a high threshold value show that it is a viable alternative to the circuit model using error-correction codes in terms of fighting against noise and error. Many physical systems have been studied to realize the MBQC, and proof-of-principle experimental demonstrations have been made, such as in photonic, continuous-variable, trapped atoms and ions, and superconducting systems. However, each system has its own challenges lying ahead that need to be overcome before a realistic one-way quantum computer can be constructed.

## Further Reading

Edge, G. J., Anderson, R., Jervis, D., McKay, D. C., Day, R., Trotzky, S., & Thywissen, J. H. (2015). Imaging and addressing of individual fermionic atoms in an optical lattice. *Physical Review A*, *92*(6), 063406.

Else, D. V., Schwarz, I., Bartlett, S. D., & Doherty, A. C. (2012). Symmetry-Protected Phases for Measurement-Based Quantum Computation. *Phys. Rev. Lett.*, *108*(24), 240505. https://doi.org/10.1103/PhysRevLett.108.240505

Farhi, E., Goldstone, J., Gutmann, S., & Sipser, M. (2000). Quantum computation by adiabatic evolution. *ArXiv Preprint Quant-Ph/0001106*.

Fenner, S. A., & Zhang, Y. (2001). Universal quantum computation with two-and three-qubit projective measurements. *ArXiv Preprint Quant-Ph/0111077*.

Feynman, R. P. (1982). Simulating Physics with Computers. *International Journal of Theoretical Physics*, *21*, 467–488.

Feynman, R. P. (1985). Quantum mechanical computers. *Optics News*, *11*(2), 11–20.

Fitzsimons, J. F. (2017). Private quantum computation: An introduction to blind quantum computing and related protocols. *Npj Quantum Information*, *3*(1), 1–11.

Fowler, A. G., Mariantoni, M., Martinis, J. M., & Cleland, A. N. (2012). Surface codes: Towards practical large-scale quantum computation. *Physical Review A*, *86*(3), 032324.

Fowler, A. G., Stephens, A. M., & Groszkowski, P. (2009). High-threshold universal quantum computation on the surface code. *Physical Review A*, *80*(5), 052312.

Frembs, M., Roberts, S., & Bartlett, S. D. (2018). Contextuality as a resource for measurement-based quantum computation beyond qubits. *New Journal of Physics*, *20*(10), 103011.

Fujii, K., & Morimae, T. (2012). Topologically protected measurement-based quantum computation on the thermal state of a nearest-neighbor two-body Hamiltonian with spin-3/2 particles. *Physical Review A*, *85*(1), 010304.

Fujii, K., Nakata, Y., Ohzeki, M., & Murao, M. (2013). Measurement-based quantum computation on symmetry breaking thermal states. *Physical Review Letters*, *110*(12), 120502.

Gao, W.-B., Yao, X.-C., Cai, J.-M., Lu, H., Xu, P., Yang, T., Lu, C.-Y., Chen, Y.-A., Chen, Z.-B., & Pan, J.-W. (2011). Experimental measurement-based quantum computing beyond the cluster-state model. *Nature Photonics*, *5*(2), 117–123.

Gimeno-Segovia, M., Rudolph, T., & Economou, S. E. (2019). Deterministic generation of large-scale entangled photonic cluster state from interacting solid state emitters. *Physical Review Letters*, *123*(7), 070501.

Gottesman, D. (1997). *Stabilizer codes and quantum error correction. Caltech Ph. D* [PhD Thesis]. Thesis, eprint: quant-ph/9705052.

Gottesman, D. (1999). *The Heisenberg representation of quantum computers. In Proceedings of the XXII International Colloquium on Group Theoretical Methods in Physics*. Cambridge, MA: International Press.

Gottesman, D., & Chuang, I. L. (1999). Demonstrating the viability of universal quantum computation using teleportation and single-qubit operations. *Nature*, *402*(6760), 390.

Gross, D, & Eisert, J. (2007). Novel schemes for measurement-based quantum computation. *Physical Review Letters*, *98*(22), 220503.

Gross, David, Eisert, J., Schuch, N., & Perez-Garcia, D. (2007). Measurement-based quantum computation beyond the one-way model. *Physical Review A*, *76*(5), 052315.

Gross, David, Flammia, S. T., & Eisert, J. (2009). Most quantum states are too entangled to be useful as computational resources. *Physical Review Letters*, *102*(19), 190501.

Gu, Z.-C., & Wen, X.-G. (2009). Tensor-entanglement-filtering renormalization approach and symmetry-protected topological order. *Physical Review B*, *80*(15), 155131.
28